# Quantum transport in double-gated graphene devices

J. Velasco Jr., Y. Lee, L. Jing, G. Liu, W. Bao, C. N. Lau

Department of Physics and Astronomy, University of California, Riverside, CA 92521

**Abstract**
Double-gated graphene devices provide an important platform for understanding electrical and optical properties of graphene. Here we present transport measurements of single layer, bilayer and trilayer graphene devices with suspended top gates. In zero magnetic fields, we observe formation of *pnp* junctions with tunable polarity and charge densities, as well as a tunable band gap in bilayer graphene and a tunable band overlap in trilayer graphene. In high magnetic fields, the devices' conductance are quantized at integer and fractional values of conductance quantum, and the data are in good agreement with a model based on edge state equilibration at *pn* interfaces.

## I. Introduction

As an electron system, graphene[1-3] has many unique properties not found in conventional semiconductor heterostructures[4-7]. For instance, the honeycomb lattice has two atoms per unit cell, giving rise to a new quantum number, pseudospin, and chiral charge carriers. Another important consequence of the honeycomb lattice is the unusual band structure: single layer graphene has a linear energy dispersion relation, thus the charge carriers behave as chiral massless particles; these particles are akin to neutrinos (but with charge *e*) or photons (but with spin ½). Moreover, due to graphene's atomic thickness, separation between adjacent sub-bands is much larger than conventional two-dimensional electron systems (2DES); thus phenomena that traditionally only emerge at dilute refrigerator temperatures become readily accessible at much higher temperatures, enabling studies that would be otherwise difficult or impossible.

An important characteristic of graphene is that, because of its zero band gap, its charge density and type are tunable via electrostatic means. One of the unique device configuration enabled by graphene is double-gated geometry, consisting of graphene positioned between a global back gate (typically provided by the doped Si substrates) and a metallic top gate. By modulating voltages applied to both gates, one can create regions with different dopant levels and types in a single graphene device, thus realizing pnp junctions with tunable junction polarity. For bilayer and trilayer grpahene, the presence of two gates also enable independent tuning of total charge density and out-of-plane electric field that breaks the inversion symmetry of few-layer graphene[8-20]. In the past few years, double-gated graphene devices have become a fascinating platform that harbors an enterprise of novel phenomena, such as Klein tunneling[21-25], Fabry-Perot oscillations[24, 26], edge state equilibration[27-30] and localization-induced conductance fluctuations[31] in the quantum Hall regime, band gap opening in bilayer[13, 14, 16, 18-

20], stacking-dependent band gap or band overlap in trilayer graphene[32-34] and observation of states with spontaneous broken symmetries in bilayer graphene[35-37].

Here we describe and discuss our transport measurements on single layer (SLG), bilayer (BLG) and trilayer (TLG) graphene devices with double-gated geometry. The article is organized as follows. Section II describes fabrication of devices with suspended top gates using a multi-level lithography technique, where the absence of dielectric material (other than vacuum) enables synthesis of high mobility devices. Section III, IV and V present and discuss transport measurements in SLG, BLG and TLG devices, respectively, in both zero and finite magnetic fields. Lastly, we conclude with a brief summary and outlook in section VI.

## II. Fabrication

Since graphene consists of only single atomic layer of carbon, deposition of top gate materials may introduce defects, dopants or additional scattering sites, which obscure precluding access to many of the most fascinating physical phenomena. Previously, materials such as electron beam resists[29] or alternating layers of $NO_2$, trimethylaluminum and $Al_2O_3$[25, 27] have been used as the dielectrics for local gates. We developed a multi-level lithography technique to fabricate 'air-bridge'-styled top gates, in which a metallic bridge is suspended across a portion of the graphene sheet, with vacuum acting as the dielectric. A similar technique was also reported by ref. [38]. Because graphene is only exposed to conventional lithographical resists and developers during fabrication, this technique minimizes the damage to the atomic layer. Also, the absence of a top gate dielectric allows post-fabrication annealing[39] to further improve device quality.

The substrate consists of a 300 nm $SiO_2$ layer grown over degenerately doped silicon that functions as a global back gate. Graphene sheets are then deposited onto the substrate by the standard micro mechanical exfoliation method, and the layer number is identified by the color contrast in an optical microscope and verified by Raman spectroscopy[40].

To fabricate the air bridge, we employ a resist bilayer with different exposure and developing conditions, so as to create a temporary support for the suspended structure; this resist support is subsequently removed after triple-angle metal depositions at −45, 45 and 0°. The details of the fabrication procedure are discussed in [26, 30]. Finally, using standard electron beam lithography, electrodes are deposited consisting of 10 nm of Ti and 80 nm of Al or Au on graphene. To ensure a robust suspension of the air bridge with none to minimal sagging, a critical point dryer is typically used during fabrication of the top gate and electrodes. As shown by the scanning electron microscope (SEM) image of a finished device in Fig. 1a, the top gate is suspended above the graphene sheet between source and drain electrodes. We note that it is important to fabricate the electrodes in the last step, since contact resistance tends to deteriorate during lithographical processes.

Since no dielectric material directly contacts the single atomic layer, these graphene devices are exceedingly clean. The air-gap between the top gate and graphene also allows surface absorbates, including those directly under the top gate, to be removed via annealing[26, 30]. A demonstration of this is shown in Fig.1b, where we measure the two-terminal conductance G of a graphene device as a function of applied back gate voltage $V_{bg}$ both before and after annealing in vacuum at 120 ºC. Clearly the $G(V_{bg})$ characteristics are considerably improved upon annealing, with a sharper Dirac point that is closer to zero, indicating the removal of undesirable absorbates and resist residue.

## III. Single Layer Graphene *pnp* Junctions

By modulating voltages applied to back gate $V_{bg}$ and top gate $V_{tg}$, a graphene *pnp* junction can be created. In a typical device, the source drain electrode separation is $L$~1-4 µm, and the width of the graphene sheet is $W$=0.4-1.5 µm; the contactless top gate straddles the center portion of the sheet with a width of ~0.5 µm, and is suspended at a distance $d$~50 - 200 nm above the substrate. Since it only covers the center portion of the graphene sheet, the top gate is *local*. Charge density in the uncovered regions, or region 1, are only controlled by voltages applied to the back gate $V_{bg}$,

$$n_1 = C_{bg}(V_{bg} - V_{bg}^D)/e, \qquad (1a)$$

whereas that in the top-gated region (or region 2) is modulated by both gates,

$$n_2 = n_1 + C_{tg}(V_{tg} - V_{tg}^D)/e. \qquad (1b)$$

Here $C_{bg}$ ($C_{tg}$) is the capacitance per unit area between graphene and the back gate (top gate), $V_{bg}^D$ and $V_{bg}^D$ are the charge neutrality points (CNP) in the "bare" and top-gated regions, respectively, and $e$ the electron charge. For a typical device on Si/SiO$_2$ substrates, the gate coupling factor $\alpha_{bg}=n_1/V_{bg}$ ~ 7.2x10$^{10}$ cm$^{-2}$. All devices are measured in a He$^3$ refrigerator at base temperature $T$=260 mK using standard lock-in techniques. The electrical lines in the refrigerator are outfitted with 3 stages of filters to ensure low noise and low electron temperature.

To investigate the device behavior in zero magnetic field, we measure its two-terminal resistance $R$ as functions of $V_{bg}$ at different values $V_{tg}$. The representative curves are shown in Fig. 1c from a device with mobility ~10,000 cm$^2$/Vs. At $n_1=n_2$=0, the device is at its global charge neutrality point, reaching a global resistance maximum $R_{max}$~7kΩ. With increasing $|V_{tg}|$, $R_{max}$ decreases to ~ 4 kΩ, and a prominent side peak develops; the portion of the $R(V_{bg})$ curves between the 2 peaks corresponds to the behavior of a bipolar (*pnp* or *npn*) junction. These features can be more easily seen in the two-dimensional plot in Fig. 1d, which plots $R$(color) as functions of $V_{bg}$ (vertical axis) and $V_{tg}$(horizontal axis). The horizontal and diagonal red lines indicate local resistance maxima and correspond to the CNP of region 1 and region 2, respectively. The slope of the diagonal line yields the ratio $C_{tg}/C_{bg}$, and is often used to determine the top gate's coupling factor $\alpha_{tg}$. Taken together, these 2 lines partition the plot into four regions with different dopant combinations– *pnp, pp'p, npn* and *nn'n*, thus demonstrating a graphene *pnp* junction with tunable junction polarity.

In high magnetic fields the cyclotron orbits of charge carriers in graphene coalesce to form Landau levels (LLs) with energies [4, 5, 7]

$$E_{SLG} = \pm\sqrt{2\hbar v_F^2 eB|N|} \qquad (2)$$

Between the LLs, the two-teminal device conductance is quantized at

$$G = 4(N+\frac{1}{2})\frac{e^2}{h} \equiv \upsilon\frac{e^2}{h} \qquad (3)$$

where $h$ is Planck's constant, $v_F$ is the Fermi velocity, $N$=...-1, 0, 1,... is an integer indicating the LL index, and $\upsilon=nh/Be$ is the filling factor. The factor of 4 denotes a degeneracy from spin and sublattice degrees of freedom. A signature of graphene's relativistic energy dispersion is the presence of a LL at $E_N=0$, which is equally occupied by electrons and holes resulting in quantum Hall plateau that are quantized at half of integer $N$ values of $4e^2/h$.

This energy spectrum of graphene is reflected in transport measurements in finite $B$. A typical data set is shown in Fig. 2a, which plots $G$ as a function of $V_{bg}$ and $B$ (the top gate is disconnected). The quantum Hall states appear as the LL fan that radiate from $B=0$ and the CNP. At $B=8T$, the device conductance $G(V_{bg})$ is quantized at 2, 6, 10, 14 and 18 $e^2/h$ (Fig. 2b), suggesting relatively small broadening of LL up to $N=4$ and indicating the high quality of the sample. We also note that the slope of the plateaus in the $V_{bg}$-$B$ plane is given by $\nu e/\alpha_{bg}h$, and can be used to extract $\alpha_{bg}$.

A local dual gated device allows regions with different doping and polarities. As a result, its quantum Hall conductance is not necessarily quantized at integer values of $e^2/h$, due to the presence of counter-propagating edge states. Instead, $G$ depends on the relative values of $n_1$ and $n_2$, and can have fractional values of $e^2/h$. A simple model was proposed by ref. [29], assuming perfect edge state equilibration at the interfaces between different regions: for a unipolar junction ($n_1n_2>0$) with $|\nu_1|\geq|\nu_2|$, the non-top-gated regions act as reflection-less contacts to the center region, yielding a device conductance

$G=e^2/h|\nu_1|$,  (4)

where $\nu_1$ and $\nu_2$ are the filling factors in the region 1 and 2, respectively. If instead $|\nu_2| > |\nu_1|$, the conductance is,

$G= e^2/h\ (1/|\nu_1| - 1/|\nu_2| + 1/|\nu_1|)^{-1}$  (5)

For a bipolar junction ($n_1n_2<0$), the device behaves simply as three resistors in series,

$G= e^2/h\ (1/|\nu_1| +1/|\nu_2| + 1/|\nu_1|)^{-1}$  (6)

The two-terminal $G$ data of a local dual gated graphene device $B=8T$ is shown in Fig. 2c as functions of $\nu_1$(vertical axis) and $\nu_2$ (horizontal axis). The most striking feature of the plot is the series of rectangles with different colors, representing quantum Hall plateaus at different combinations of $\nu_1$ and $\nu_2$. The robust appearance of the various conductance plateau up to an orbital level of $N=2$ underscores the high quality of this sample. Line traces $G(\nu_2)$ at $\nu_1=-2, 2, 6$ and $10$ are displayed in Fig. 2d, with conductance values in excellent agreement with those calculated by Eq.s (4-6). For instance, for $\nu_1=2$, $G=0.67$ at $\nu_2=-2$, $G=2$ at $\nu_2=2$, and $G=1.2$ at $\nu_2=6$. Thus our data demonstrate partial and full equilibration of co- and counter-propagating edge states.

## IV. Bilayer Graphene *pnp* Junctions

The band structure of a pristine bilayer graphene is parabolic, $E=\dfrac{\hbar^2 k^2}{2m^*}$, where $m^*=0.02$-$0.04\ m_e$ and $m_e$ is electron's rest mass, with the conduction and valence bands touching at the inequivalent $K$ and $K'$ points[8-11]. Similar to SLG, applying $V_{bg}$ and $V_{tg}$ creates a *pnp* junction in bilayer graphene with modulation of junction polarity and dopant level. A typical data set is shown in Fig. 3a, displaying two-terminal $R$ (color in a logarithmic scale) as functions of $V_{bg}$ (vertical axis) and $V_{tg}$(horizontal axis) for a device with electron mobility ~10,000 $cm^2/Vs$. In this device, the top gate straddles the center portion of the flake at a distance $d\sim50$ nm above with a width of 0.5 μm.

It is instructive to compare this data set with the SLG data in Fig. 1c. Both plots are partitioned into 4 quadrants of different junction polarities, with the diagonal line that indicate the resistance maximum $R_{max}$ at the CNP of region 2. The striking difference between the plots is the change in *magnitude* of $R_{max}$ – in SLG $R_{max}$ varies by only ~ 50%, while in BLG, $R_{max}$ changes by more than a factor of 100, reaching as high as 1 MΩ (Fig. 3b).

This unusual behavior of BLG *pnp* junction arises from opening of a band gap in the BLG energy spectrum[8, 11-20]. What sets dual-gated BLG from their SLG counterparts is that application of voltages to the two gates not only controls densities $n_1$ and $n_2$, but also an electric field $E_2$ applied across bilayer in the top-gated region. This electric field is given by

$$E_2=[\varepsilon_{sio}(V_{bg}-V_{bg}^D)/t - (V_{tg}-V_{tg}^D)/d]/2 \quad (7)$$

where $\varepsilon_{sio}$ ~ 3.9 and $t=300nm$ are the dielectric constant and thickness of the SiO$_2$ layer, respectively. Such an out-of-plane electric field breaks the BLG's inversion symmetry and creates a potential difference between the two layers, $V_2=E_2*(3.3Å)$. As a result, the band structure of BLG adopts the Mexican hat profile, with an induced gap

$$\Delta = \frac{t_\perp eV_2}{\sqrt{t_\perp^2 + e^2V_2^2}}$$ that scales linearly with $V_2$ for small $V_2$, and saturates at $t_\perp$ when $V_2$ is

sufficiently large. Here $t_\perp$ ~ 0.2-0.4 eV is the inter-layer hopping energy.

The highly resistive state observed in the upper left corner of Fig. 3a thus arises from the gap opening at large $E_2$. On the other hand, though $E_2$ as high as 1V/nm is applied, the device never become truly insulating (*e.g.* resistance does not exceed 1 MΩ), as one would expect from the opening of a gap as large as 0.2 eV. This discrepancy is due to the presence of localized states that either lie within the gap[41-44] or are formed from disorder-induced charge puddles[45, 46], and transport across the junction has been shown to be variable range hopping at low temperatures and thermal activation at high temperature[14, 17, 18, 20].

In high magnetic fields, LLs in BLG have energies

$$E^{BL} = \pm\frac{\hbar eB}{m*}\sqrt{N(N-1)} \quad (8)$$

Between the LLs, the device conductance is quantized at

$$G=4N(e^2/h) \quad (9)$$

where the LL index N=… -3,-2, -1, 1, 2, 3…. Notably, the $N=0$ and $N=1$ Landau levels are degenerate, resulting in the 8-fold degeneracy at $E=0$.

The quantum Hall conductance of a BLG *pnp* junction at $B=8T$ is shown in Fig. 3c. Just as the case for SLG, the presence of regions with different densities gives rise to co- or counter-propagating edge states, which equilibrate at the junction interfaces. This results in regions with different conductance values, appearing as bands of colors in Fig. 3c. The conductance values are also given by Eq.s (4-7), with one important difference – instead of -6, -2, 2, 6…, the filling factors $v_1$ and $v_2$ are -8, -4, 4, 8...*etc.* As shown in Fig. 3d, line traces $G(v_2)$ at $v_1$=4, 8, and 12 are in reasonable agreement with theoretical calculations[20].

Finally, we note that in the absence of broken degeneracy, the smallest quantum Hall conductance of BLG is 4 $e^2/h$ for a uniform junction, and 4/3$e^2/h$ for a *pnp* junction. In contrast, an insulating state appears in Fig. 3c, as represented by the dark blue region.

This insulating state emerges at high $B$ and $E_2$, and is an indication of the competing symmetries in BLG. It is also intimately related to the intrinsic insulating state in ultra-clean BLG[36, 37], the nature of which has been a topic of intense debate to date[47-56].

## V. Trilayer Graphene *pnp* Junctions

Trilayer graphene has two stable configurations, ABA and ABC stackings, which differ only in the relative positions of the topmost and bottom layers. As the name suggests, in ABA-stacked TLG the top and bottom layers are AA-stacked, whereas in ABC-stacked TLG is they are Bernal-stacked. These 2 allotropes have dramatically different band structures[32, 57-62] and electrical and optical properties[32-34, 63-66], and can be experimentally identified via infrared or Raman spectroscopy[63, 67].

Here we will focus on only ABA-stacked TLG, which is estimated to consist of ~85% of all graphene devices. Its band structure is expected to be a combination of SLG's linear dispersion and BLG's quadratic dispersion. Like BLG, it is susceptible to effects of an out-of-plane electric field $E_2$; but instead of the tunable band *gap* in BLG, ABA-stacked TLG is predicted to have a tunable band *overlap*[32, 57-62]. Such effects can be seen from transport data of double-gated TLG device, which is verified to be ABA-stacked via Raman spectroscopy (Fig. 4a). Fig. 4b displays the resistance $R$ (color) through a TLG *pnp* device with suspended top gates as functions of $V_{bg}$ (vertical axis) and $V_{tg}$ (horizontal axis). The global Dirac point of the device is found to be at $V_{bg}$=-7.8 and $V_{tg}$=13.5V. Fig.4c displays the line traces $R(V_{bg})$ at different $V_{tg}$ values. For $V_{tg}$ at the Dirac point, the $R(V_{bg})$ curve has the characteristic inverse V-shape, with an estimated mobility 4000cm$^2$/Vs. As $V_{tg}$ shifts away from the Dirac point, *i.e.* with increasing $E_2$, the maximum resistance $R_{max}$ decreases, indicating an increasing band overlap, in agreement with previously reported work[32].

In magnetic fields, from tight-binding calculations that keep only the nearest-layer coupling term, the Landau level spectrum for ABA-stacked trilayer[9, 57, 68, 69] is a superposition of those for SLG and BLG, *i.e.*, a combination of Eq. (2) and (8). The LL at zero energy is 12-fold degenerate, giving rise to quantized conductance plateaus with integer values ...-10, -6, 6, 10, 14... of $e^2/h$. When other interlayer coupling terms are included ($\gamma_2$-$\gamma_5$ in the Slonczewski-Weiss-McClure parametrization of graphite), some of the degeneracies are expected to be broken. However, the LL at zero energy is expected to retain at least 2-fold degeneracy.

Fig. 4d displays $G$ (in units of $e^2/h$) of the TLG device at $B$=8T as functions of $V_{bg}$ (vertical axis) and $V_{tg}$ (horizontal axis). Notably, an insulating state $\nu$=0 was observed, indicating partial lifting of the 12-fold degeneracy. We note that this is the first time that the insulating state is observed in ABA-stacked TLG, and realization of a *pnp* junction in a trilayer device. To compare with SLG and BLG devices in the quantum Hall regime, we replot the data in Fig. 4d as functions of $\nu_1$ and $\nu_2$ (Fig. 4e), which resembles Fig. 2c and Fig. 3c. Line traces $G(\nu_2)$ at $\nu_1$=-6 and -10 are shown in Fig. 4f. We find that the conductance values are only in semi-quantitative agreement with Eq.s (4-6). For instance, for $\nu_1$=-10, $G$ is expected to reach 10$e^2/h$ at $\nu_2$=-10 and decreasing to ~7.8 $e^2/h$ at $\nu_2$=-14 and ~6.9$e^2/h$ at $\nu_2$=-18. From the blue trace in Fig. 1f, $G$~8.5$e^2/h$ at $\nu_2$=-10, and ~6.9$e^2/h$ at $\nu_2$=-18, but with no discernible plateau at $\nu_2$=-14. Such discrepancy between the model and data is likely due to the relatively low mobility of the TLG device, and is expected to be diminished or eliminated with improved device quality.

## VI. Conclusion

Graphene and its few layer cousins constitute a unique electron system, and double-gated graphene devices have been a particularly attractive platform for investigation of electric properties in zero magnetic field and in the quantum Hall regime. An important frontier in this area is ultra-clean double-gated graphene devices, which can be fabricated by suspending graphene[35-37] or use BN as the top and bottom gate dielectrics[70, 71]. Many novel phenomena, both predicted and unforeseen, are expected to emerge in these ultra-clean systems. Examples include possible phase transitions driven by electric and magnetic fields in BLG and TLG, interactions among edge states with broken degeneracies and in ballistic devices, and Klein tunneling in few layer graphene. Thus, we expect double-gated graphene devices continue to offer much insight into low dimensional quantum systems.

**Acknowledgement**
The authors acknowledge the support by NSF CAREER DMR/0748910, NSF DMR/1106358, NSF ECCS/0926056, ONR N00014-09-1-0724, ONR/DMEA H94003-10-2-1003 and the FENA Focus Center. CNL acknowledges the support by the "Physics of Graphene" program at KITP.

Fig. 1. Device image and data of SLG at $B=0$. (a). SEM image of a graphene device with a suspended top gate. (b). $G(V_{bg})$ of a typical graphene graphene before (blue line) and after (red line) annealing. (c). $R(V_{bg}, V_{tg})$ of a double-gated SLG device. (d). Line traces through (c) at, from purple to red, $V_{tg}$=-30, -20, -10, 0, 10, 20, 29.5V, respectively.

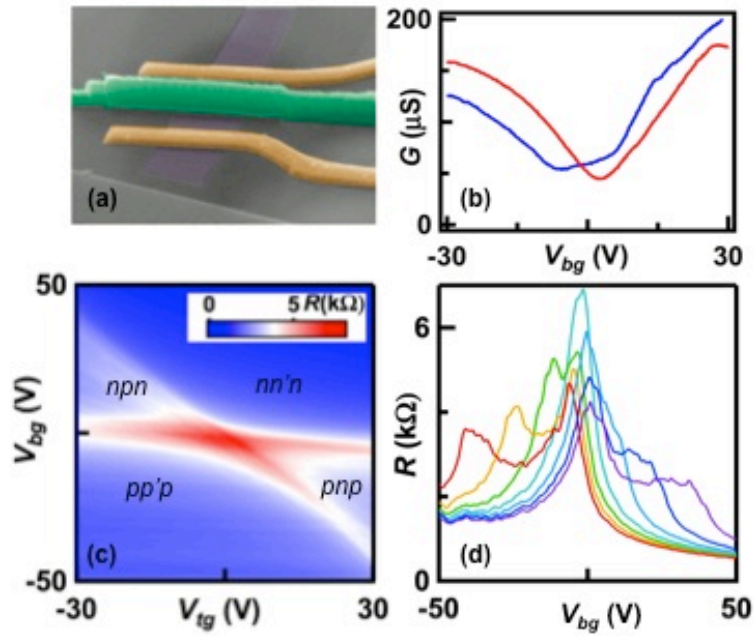

Fig. 2. Data from a SLG pnp junction in the quantum Hall regime. (a). $G(V_{bg}, B)$ of a SLG device. (b). $G(V_{bg})$ of the device at $B=8$T. (c). $G(v_1, v_2)$ of a SLG at $B=8$T. (d). Line traces $G(v_2)$ at $v_1=-2, 2, 6$ and $10$.

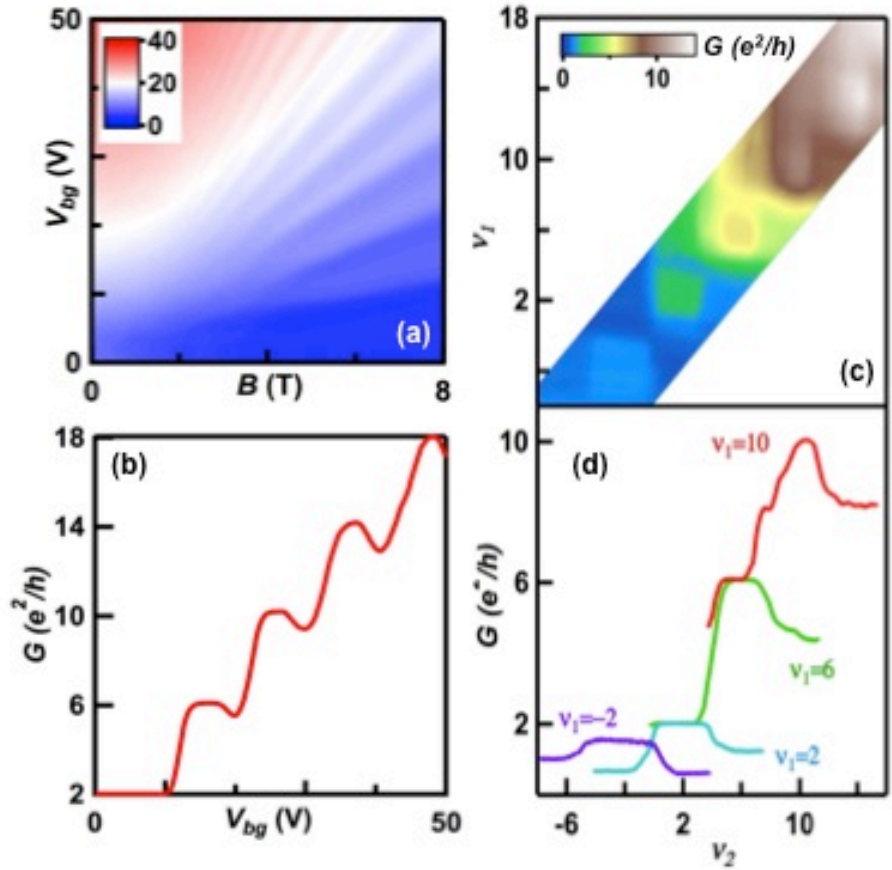

Fig. 3. Transport data from a BLG *pnp* junction. (a). $R(V_{bg}, V_{tg})$ of the BLG device at $B=0$. Note that the logarithmic color scale. (b). Line traces through (a) at, from red to purple, $V_{tg}$=-30, -20, -10, 0, 10 and 15V, respectively. (c). $G(v_1, v_2)$ of the device at $B=8T$. (d). Line traces $G(v_2)$ at $v_1$=4, 8 and 12.

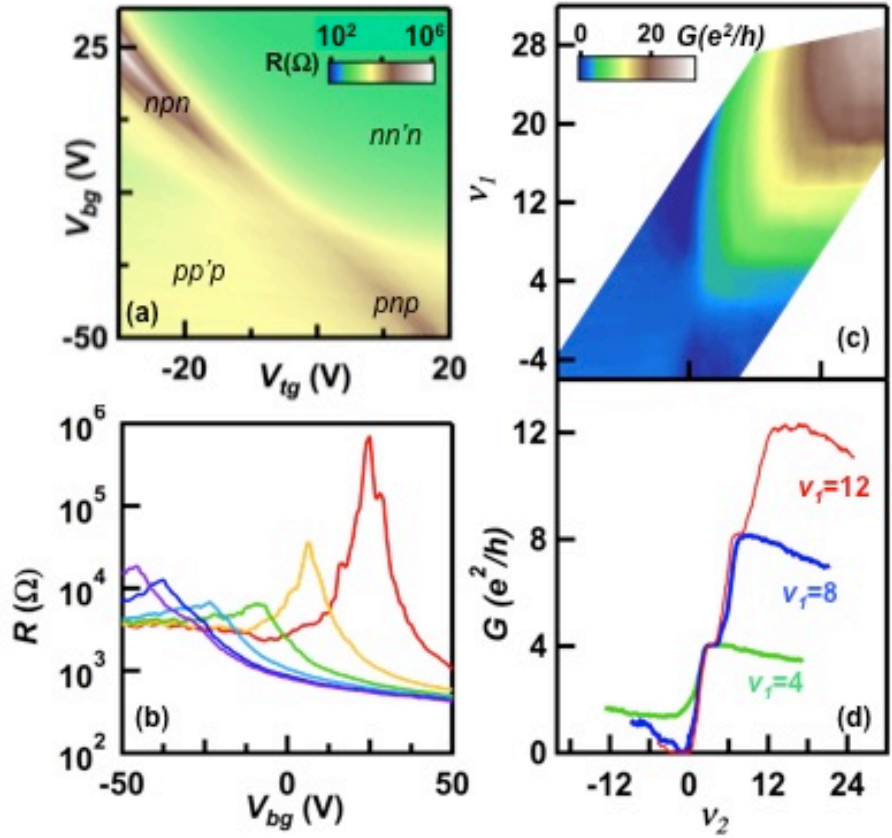

Fig. 4. Raman and transport data from a TLG *pnp* junction. (a). Raman spectrum of the TLG device. (b). $R(V_{bg}, V_{tg})$ in k$\Omega$ of the TLG device at $B=0$. (c). Line traces through (b) at, from red to blue, $V_{tg}$=13.5 (CNP), 0, 10, 20, and 28V, respectively. (d). $G(V_{bg}, V_{tg})$ in units of $e^2/h$ of the TLG device at $B=8$T. (e). Data in (d) replotted as functions of $v_1$ and $v_2$. (d). Line traces $G(v_2)$ at $v_1$=-6 (red) and -10 (blue).

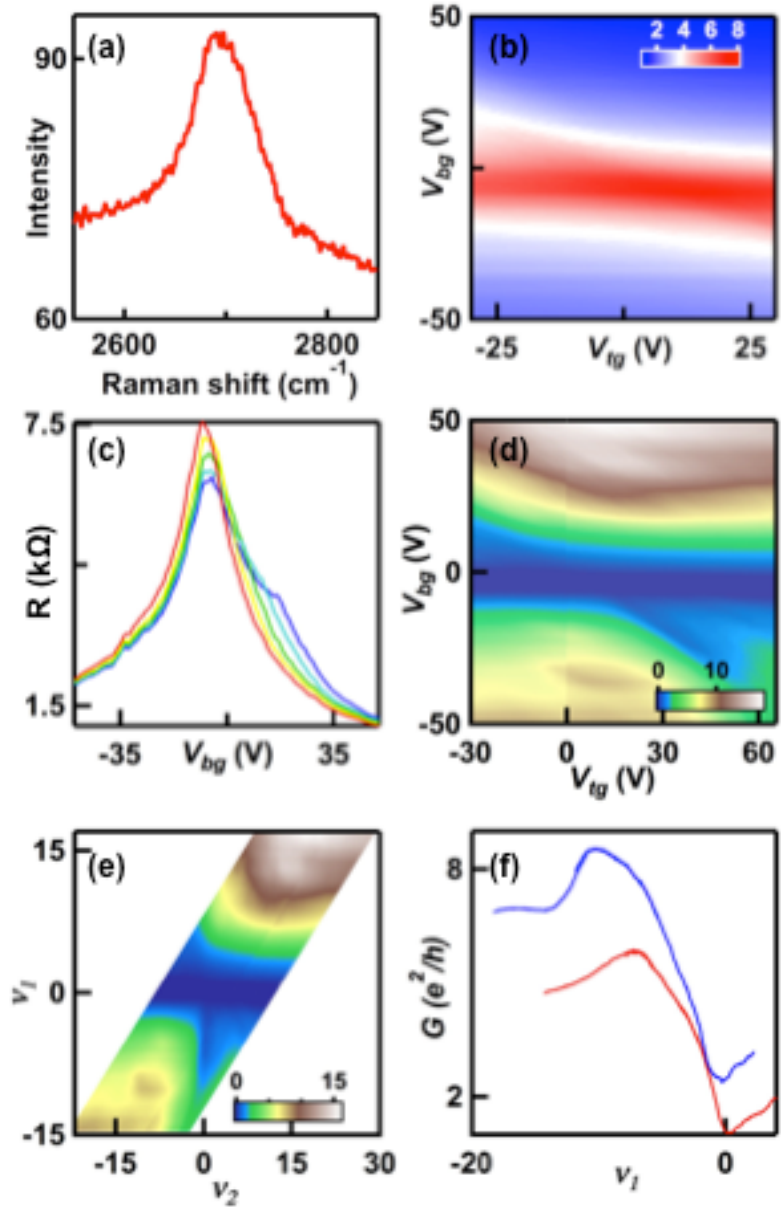